\documentclass[runningheads]{llncs}
\usepackage[T1]{fontenc}
\usepackage{subcaption}
\usepackage{graphicx}
\begin{document}
\title{Understanding emotional and behavioural responses through physiological signal analysis recorded during theatrical performance}
%
%
\author{Aditi Site\inst{1}\orcidID{0000-0001-9802-2061} \and
Annariina Lohiranta\inst{1} \and
Tarmo Lipping\inst{1}\orcidID{0000-0002-5112-2425}}
\authorrunning{F. Author et al.}
%
\institute{Tampere University, University consortium of Pori, Pohjoisranta 11A, 28100 Pori
\email{name.surname@tuni.fi}}
\maketitle              
\begin{abstract}

Among various cultural interventions, theatre offers a powerful medium for emotional expression, reflection, and social connection. Live performances can positively influence mental well-being. Therefore, studying audience responses during theatrical performances can provide valuable insights into the relationship between theatre, emotional experiences, and well-being. Physiological responses have emerged as an important modality for identifying patterns associated with different emotional states. However, in naturalistic environments such as live performances, analysing these signals is challenging due to the absence of explicit labels. In this study, we focus on understanding emotional and behavioural responses using physiological data collected during a live theatrical performance. Data were recorded from participants using multiple sensing modalities. In addition, a comprehensive analytical framework was employed, which included demographic analysis, visual exploration of physiological signals, participant-level observations, scene-wise physiological analysis, and intensity-based analysis using z-score normalization to summarize the findings. The findings revealed that dramatic, surprising, and intense scenes produced the strongest activation, often with concurrent increases in HR and SCL, whereas empathy-driven scenes showed lower and more stable responses. Humorous scenes elicited moderate-to-high activation, particularly in the presence of sudden reactions or loud sounds. Analyses also identified shared physiological response patterns across participants, with activation concentrated around key emotional events. These results indicate that combined HR and SCL measures can effectively capture variations in audience engagement during live theatrical performances.

\end{abstract}
\section{Introduction and Research Background}
Mental and social well-being are important aspects of overall health and demonstrate strong links to physical health. Cultural interventions such as dance therapy, art therapy, music, and theatre, movies have gained increasing attention for their ability to promote emotional expression, social connectedness, and psychological well-being \cite{Theatrestudy0,TheatreStudyBackground}. Especially stories have a ability to capture attention, evoke emotions, and immerse audiences in the experiences of their characters. Through narrative progression, audiences often become emotionally invested in the relationships and development of characters, allowing them to experience a range of emotions throughout the story. This quality is particularly pronounced in live or theatrical performances, where the physical presence of actors, real-time interactions, and shared audience experience create a strong sense of engagement \cite{Theatrestudy1}. In addition to this, there are several other cultural activities that contribute to emotional and social well-being, including music-based interventions, dance and movement therapy, visual arts practices, festival gathering, storytelling sessions, and community-based participatory art programs \cite{Theatrestudy0,Theatrestudy2}.

Recently, a wide range of techniques have been developed to measure emotional experiences and, consequently overall well-being \cite{Wellbeingmonitoring}. These include self-report questionnaires, behavioral observation, and increasingly, analysis of physiological signals such as Electrocardiogram (ECG), Galvanic Skin response (GSR), Photoplethysmogram (PPG) and Electroencephalogram (EEG) \cite{TheatrestudySubAnalysis,PhysioSigbyMusic,Theatrestudy0}. These methods are able to capture dynamic emotional states in real-world settings, providing deeper insights into how individuals experience and respond to different environments and interventions. Emotional experiences are commonly characterized using discrete emotion categories such as happiness, sadness, anger, fear, disgust, and surprise as well as dimensional models such as valence and arousal \cite{Russell_circumplex_model}. Cardiovascular measures such as HR are commonly associated with changes in autonomic activation and emotional arousal, while EDA/SCL is particularly sensitive to sympathetic activation and changes in arousal. Study in \cite{CharacterizingEmotions0} shows how different emotions maps to the body. It also indicates how heart rate changes in case of empathy, sadness and fear kind of situations. However, physiological responses do not provide a unique physiological signature for every discrete emotion and are influenced by emotional intensity and contextual factors. Therefore, HR and SCL are generally considered indicators of physiological activation rather than direct measures of specific emotions. Studies in \cite{CharacterizingEmotions1} suggests that high-arousal emotions such as fear and anger have frequently been associated with increases in heart rate (HR) and skin conductance level (SCL), reflecting increased autonomic activation whereas physiological responses to positive emotions such as amusement or joy appear to be more variable and dependent on the stimulus and context.

Many studies have used physiological signals to analyze emotional and engagement responses in various setting such as music listening, theatrical performances, ballet events, robot assisted interventions \cite{Robotassistedrecreation,Robotassistedrecreation1}, and extracurricular activities targeting students \cite{Extracurricularrecreation}. Furthermore, significant developments have been observed in emotion recognition using both supervised and unsupervised machine learning techniques using various physiological signals \cite{UnsupervisedMLforphysioRecog,RealTimePsychowellbeing,EmoRecoModels,EmoRecoModels1,EmoRecoModels2}. Researchers in \cite{PersonlaizedEmoRecoModels,PersonlaizedEmoRecoModels1} have used personality traits along with physiological signals to train emotion recognition system. These approaches leverage physiological and behavioral data to identify patterns associated with different affective states. Such advancements have the potential to enhance the development of automated affective state analysis systems, supporting real-time detection and interpretation of human emotions in naturalistic settings \cite{WellbeinginNaturalEnvironment}.

Following this line of research, the objective of this study is to investigate the patterns in the physiological responses that are associated with different emotions during a live performance without relying on explicit labels. This study examines whether common physiological response patterns emerge across participants during key scenes of the performance. In this direction this study aims to explore the following research questions:
\begin{enumerate}
    \item How do physiological signals vary during different emotional scenes in a live theatrical performance?
    \item How consistent are physiological responses across participants during key theatrical moments?
    \item What differences can be observed between different parts of the performance in terms of overall physiological intensity?
    \item  How effectively can z-score–based intensity analysis and visualization techniques capture arousal/engagement patterns in the participant?
\end{enumerate}

\section{Data Collection and Experimental Setup}
This study was conducted to examine emotional and behavioral responses during theatrical performances through the analysis of physiological signals, with the objective of identifying underlying patterns and correlations between physiological activity and emotional states.
The experimental study was designed to capture real-time audience reactions to different theatrical situations and emotional stimuli presented during the performance. The study was approved 
by the Human Sciences Ethics Committee of the Universities 
in Satakunta, Finland (approval ID: 16.05.2025). The following sections provide details regarding the participants, experimental setup, data collection procedure, physiological signals acquired, and the key scenes selected for analysis.
\subsection{Theatrical Performance Overview}
The experimental study was conducted during a live theatrical performance. The performance comprised of multiple scenes with varying sound and lighting effects, emotionally intense moments, and sequences intended to evoke amusement, empathy, and suspense. In here, "empathy" was used to describe scenes containing situations that were expected to evoke empathic or sympathetic responses from the audience, particularly scenes involving sadness, concern, pity. The performance lasted approximately 125–130 minutes and was divided into two parts. The first part had a duration of approximately 80 minutes, followed by a 20-minute of break. The second part lasted approximately 45 minutes, completing the overall performance. 

The theatrical performance used in this study, named \textbf{"Isin tyttö (Daddy’s Girl)"}, is a cutting-edge black comedy and thriller about family dynamics and survival. Several key scenes within the theatrical performance were identified that represent critical moments in the storyline where changes in tone, intensity, or audience engagement are expected. Furthermore, some of these key scenes correspond to flashing lights and loud sounds. These key scenes were selected to capture a range of emotional and narrative dynamics, enabling a structured analysis of audience’s physiological responses across different phases of the performance. Table \ref{tab:keyscenes} summarizes the key events of the theatrical performance, including the part, scene reference, description, and expected emotion. In the subsequent sections, the scenes are referenced consistently in the same order as presented in the table.

\begin{table}
\caption{Key events in the theatrical performance}
\label{tab:keyscenes}
\begin{tabular}{|p{1cm}|p{1.5cm}|p{7cm}|p{3cm}|}
\hline
\textbf{Part} & \textbf{Scene} &  \textbf{Description}  & \textbf{Expected emotion}\\
\hline
Part 1 & Scene 1  &  The main character (daughter) receives and responds to birthday gift & Amusement\\
 & Scene 2  & The main character (daughter) loses a tooth during the performance, which is portrayed comically  & Amusement\\
 & Scene 3  & The main character gets punishment & Empathy \\
 & Scene 4  & Intense discussion between the main character's parents  & Intense \\
 & Scene 5  & Family Olympic-style games at a summer cottage with loud cheering and playful interaction  & Amusement \\
 & Scene 6  & Father establishes strict household rules during family meeting & Amusement\\
 & Scene 7  &  Father reacts angrily for mocking his auction and throws objects in an emotional outburst & Sudden reaction/ Emotional arousal\\
\hline
Part 2 & Scene 1  & Family Olympic-style game where the main character wins and expresses joy, presented in a comic manner & Amusement\\
 & Scene 2  & The main character's efforts to become sick are portrayed comically through exaggerated actions and audiovisual effects & Amusement\\
 & Scene 3  & Intense father–daughter argument followed by a symbolic shooting portrayed on stage & Surprise/Intense reaction\\
 & Scene 4  & Father creates an uncomfortable sauna situation for the daughter's boyfriend in a humorous manner & Amusement\\
 & Scene 5  & Family members share jokes about the father & Amusement\\
 & Scene 6  & Father reacts angrily to missing salt in the food with an abrupt emotional outburst & Sudden reaction\\
 & Scene 7  & Father is hit by a wall during the scene & Surprise/ Sudden reaction\\
 & Scene 8  & Mother shares a personal story about her relationship with the father & Empathy\\
\hline
\end{tabular}
\end{table}

\subsection{Participant Demographics}
A total of 17 participants were recruited for the study to record physiological responses during the theatrical performance. The sample included participants of varying age groups and genders to ensure diversity in the recorded responses. Participants were recruited and briefed about the procedure prior to data collection. The criteria for participation was that the person is over the age of 18 and has no diagnosed long-term illnesses that affect the functioning of the central nervous system. Theatre groups were responsible for selection of participants and research group was responsible for sending the consent forms prior to the recording day. All participants provided informed consent prior to their inclusion in the study. Basic demographic information, including age and gender, was collected for analysis purposes. No participants reported any conditions that could significantly affect physiological signal recording.

\begin{figure}
    \centering
    \includegraphics[width=\textwidth]{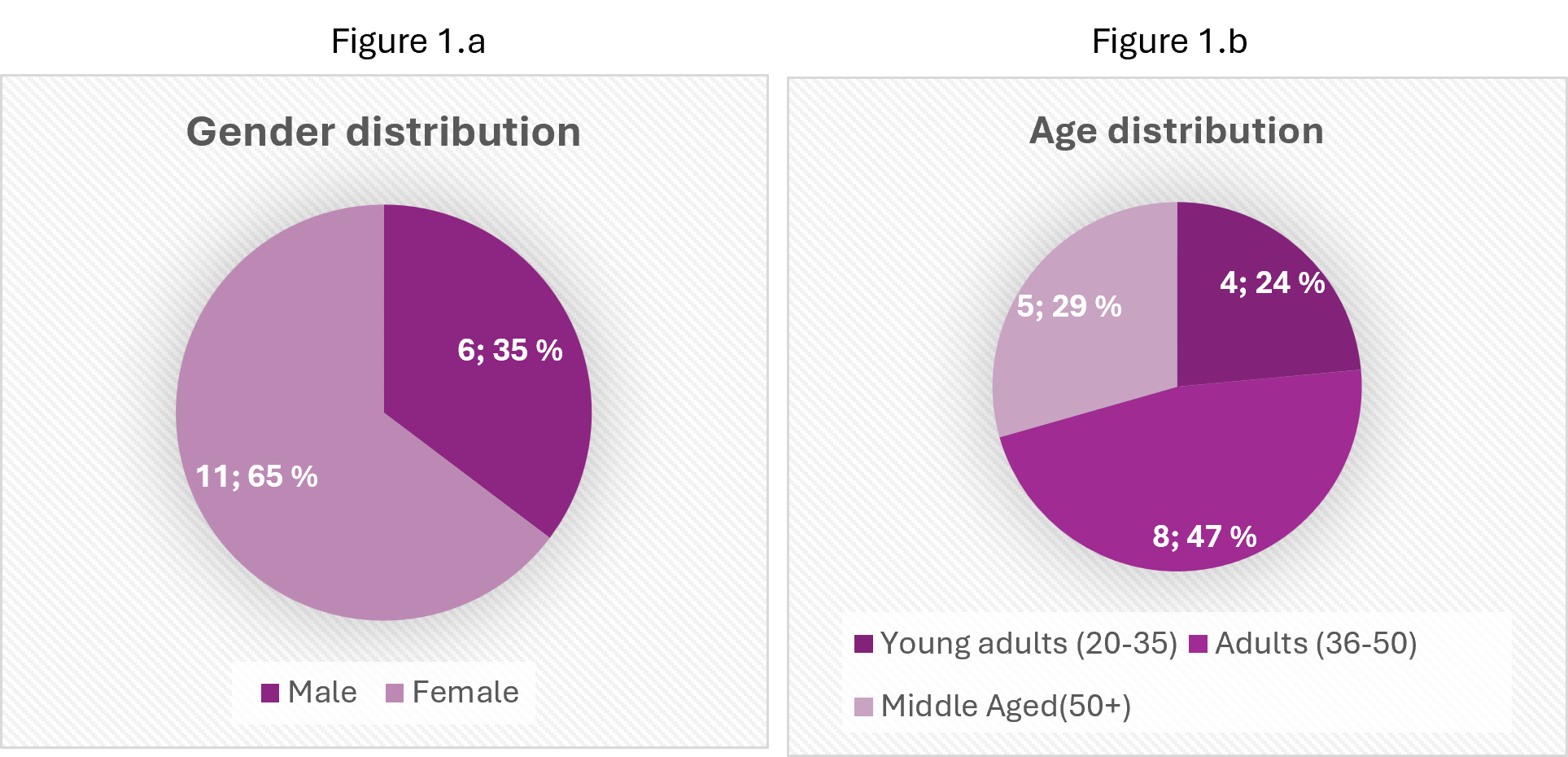}
    \caption{Demographic summary} \label{fig:Demographic summary}
    \end{figure}

The demographic distribution of the participants is summarized in Figure \ref{fig:Demographic summary}. The sample comprised 35\% male and 65\% female participants (Figure 1.a). Participants ranged in age from 20 to 68 years, with the majority of them between 36–50 years (Figure 1.b). The cohort included all participants with Finnish nationality. Overall, the sample reflected a diverse demographic distribution suitable for the study objectives.

\subsection{Data Collection Procedure}
The data collection was conducted during a live theatrical performance in a naturalistic audience setting. Participants were seated in the top row to avoid any disturbances to other audience members due to measurement setup. Physiological signals were collected using Electrocardiogram (ECG) and Galvanic Skin Response (GSR). All sensors were non-invasive and attached before the performance started to ensure participant comfort and data reliability. Data were recorded continuously throughout the theatre performance to capture real-time physiological responses. In addition, video recordings of each performance were taken and time-synchronized with the physiological data to enable precise alignment of participant’s physiological responses with specific scenes and events. Since the performance took place in two parts including a break in between, the sensors were removed for the 20 minutes break time and also the video recording was paused for the duration of break. The following sensors and measurement devices were used for this study:
\begin{enumerate}
    \item \textbf{Electrocardiography (ECG)} was used to measure heart activity, particularly heart rate (HR) and heart rate variability, which are indicators of emotional arousal. ECG was collected using Shimmer sensing device using 3 electrodes. 
    \item \textbf{Galvanic Skin Response (GSR)} sensors were employed to assess changes in skin conductance associated with sweating, providing insights into emotional arousal and autonomic nervous system activity. For GSR also, Shimmer sensing device was used, and electrodes were places on fingers.
\end{enumerate}

\section{Methodology} \label{sec:methodology}
\begin{figure}
\centering
\includegraphics[width=\textwidth]{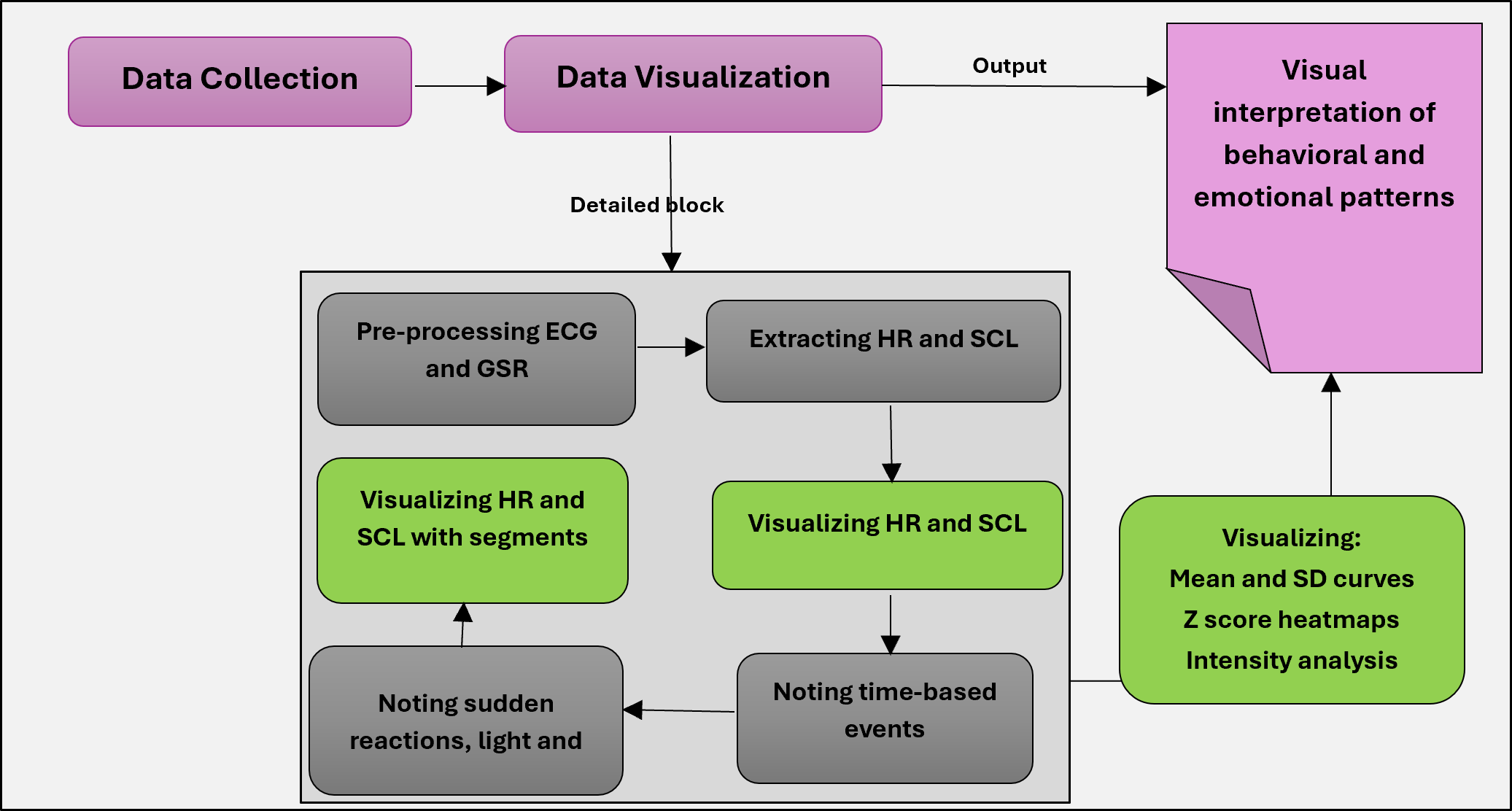}
\caption{Analytical framework} \label{fig:Analytical framework}
\end{figure}

This section describes the overall experimental methodology adopted in this study to conduct the preliminary analysis of the processed physiological signals. This section explains the framework and the visualization approach used to process and interpret the physiological signals collected during the theatrical performance. For the scope of this study we have used ECG and GSR signals for our analysis. First, the recorded physiological data were trimmed according to the exact duration of the theatrical performance. The timestamp information was converted into seconds to provide a common temporal reference for synchronization with the performance scenes. For ECG processing, the NeuroKit2 library was used \cite{Neurokit2}. This procedure includes ECG signal cleaning and R-peak detection, from which the inter-beat intervals and heart rate were derived. The resulting HR signal was then smoothed using a rolling window before further participant-level and scene-wise analysis. For EDA processing also we used NeuroKit2 library  to clean and decompose the electrodermal signal into its tonic and phasic components. The tonic component represents the slower-varying level of electrodermal activity, while the phasic component represents more rapid transient changes in skin conductance. In this study, the tonic component was used as the skin conductance level (SCL) measure because our analysis focused on overall changes in physiological activation across the performance.
The extracted HR and SCL were used to enable quantitative analysis of emotional responses across different theatrical events. Figure \ref{fig:Analytical framework} describes the methodology used for processing raw ECG and GSR signals and the green-highlighted blocks denote the visualization methods used to perform preliminary data exploration and identify patterns in the recorded physiological signals. Various visualization methods adopted in this study are described below.

\begin{enumerate}
    \item \textbf{Visual Exploration of HR and SCL Signals:}
    A visual analysis of the HR and SCL signals was performed as to support the identification of physiological arousal patterns. The temporally aligned and aggregated HR and SCL signals were plotted across the duration of the theatrical performance to facilitate qualitative inspection of participant responses over parts of performance. This visualization step was used to identify segments of elevated physiological activity, including peaks and sustained increases in HR and SCL, which are indicative of heightened arousal states. These visually detected segments were then used as reference points for identifying the key scenes and events for further analysis.
    
    \item \textbf{Event-wise visualization:}
    To identify key moments within the performance, a manual segmentation approach was followed. To improve the reliability on the scene segmentation process, two researchers who had also watched the performance reviewed the performance and were involved in identifying and segmenting the key scenes and assigning their expected emotional contexts. The individual participants may experience the same scene differently, however, continuously capturing and labeling each participant's subjective emotional experience in a real-world theatre setting was challenging without interrupting the natural viewing experience. Therefore, we considered carefully observed, researcher-based scene segmentation to be the most practical approach for the present exploratory study. First, the overall heart rate and skin conductance signals were visualized to detect periods of elevated physiological arousal. These periods were then cross-referenced with the time-synchronized video recordings of the performance. By carefully reviewing the videos, the onset and offset times of potentially significant events were manually annotated and recorded. The annotated timestamps were subsequently used to segment the physiological data into event-specific intervals. These segments were then plotted and analysed to examine participant responses associated with scenes and performance events.
    
    \item \textbf{Intensity analysis across scenes:}
     To quantify overall physiological intensity during the theatre performance, an intensity index was computed by combining HR and SCL responses. Both HR and SCL were first transformed into z-scores to standardize the data across participants. The z-score represents how much a given value deviates from an individual’s baseline mean. In this context, it captures whether physiological responses are above or below each participant’s typical resting level, making it suitable for assessing arousal. Higher positive z-scores indicate stronger deviation from baseline and therefore higher physiological arousal, while lower or negative values indicate reduced arousal relative to baseline. This normalization allows meaningful comparison across participants with different baseline physiological levels. The final intensity measure was computed as the average of HR and SCL z-scores for each time segment, providing a unified representation of autonomic arousal.
\end{enumerate}
The synchronized video and signals enabled segmentation of the performance into meaningful narrative events. Patterns of peak arousal and engagement were identified and correlated with performance moments observed in the video. This multimodal integration allowed interpretation of participant responses in both physiological and contextual dimensions. The framework provided a systematic approach to understanding how theatrical elements influence emotional and cognitive states in real time.

\section{Visual Analysis and Key Inferences}
This section presents a visual analysis of the HR and SCL signals recorded during the theatrical performance. The objective is to identify patterns in the HR and SCL associated with varying emotions throughout the performance and during key events/scenes described in Table \ref{tab:keyscenes}. By correlating HR and SCL measures with time-aligned video recordings, the study examines how specific theatrical moments influenced participant arousal, attention, and engagement.

\begin{figure}
\centering
\includegraphics[width=0.8\textwidth]{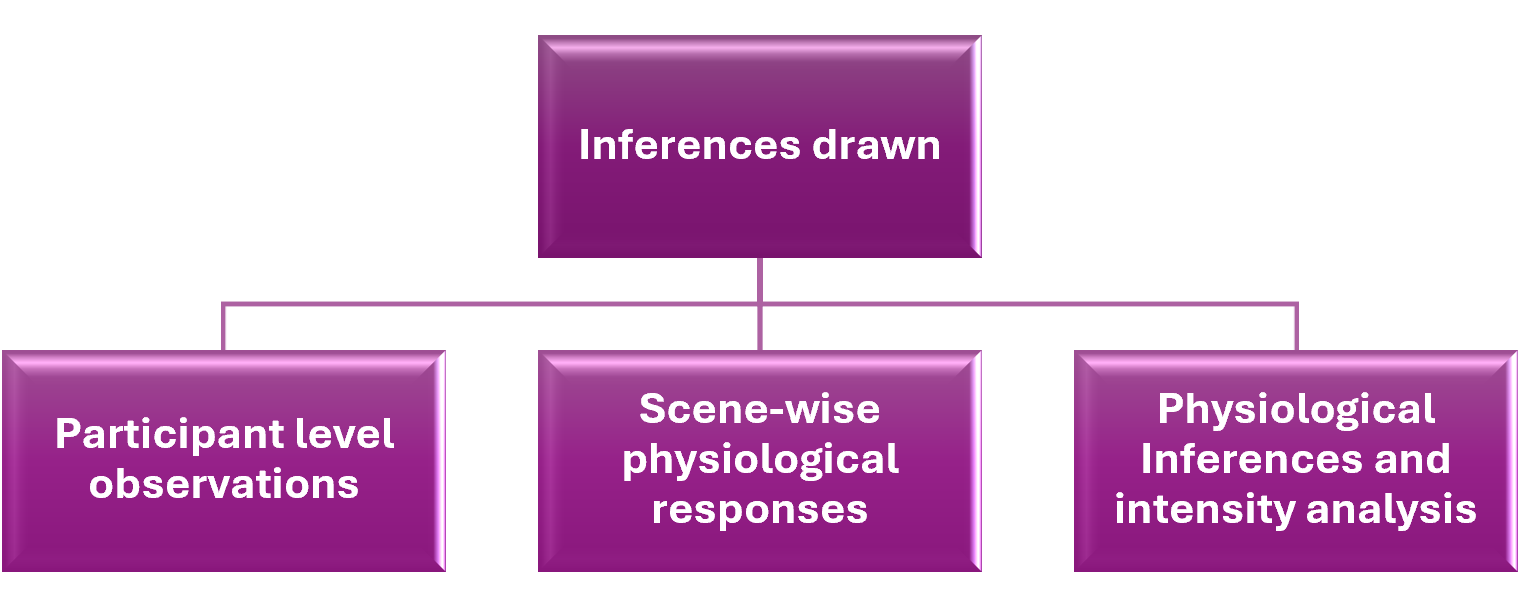}
\caption{Categories of inferences} \label{fig:Categories of inferences}
\end{figure}

The inferences are derived from three perspectives, divided in three sections as shown in Figure \ref{fig:Categories of inferences}. First, participant-level analysis is conducted to examine individual differences and overall physiological response patterns across audience members. Second, a scene-wise physiological response analysis is performed by relating variations in HR and SCL to the annotated key scenes of the performance. Third, a physiological signal-based intensity analysis is performed using a range of visualization techniques, including mean ± standard deviation (SD) curves, heatmaps, and intensity plots derived from the HR and SCL values. These visualizations are used to examine the consistency of physiological responses across participants, enabling the identification of periods of heightened arousal, prominent response peaks, and sustained activation throughout the performance.
\subsection{Participant-level observation}
To investigate individual differences and also similarities within the participants in physiological responses, the HR and SCL signals of each participant were visually examined throughout the theatrical performance. In addition to individual response patterns, this level of analysis also highlights similarities and shared trends across participants throughout the performance. The methodology used in section \ref{sec:methodology} as described in Figure \ref{fig:Analytical framework}, is used to make individual participant level plots for HR and SCL. These plots are analyzed to extract key observations and infer physiological response patterns. 

It was observed that approximately half of the participants (8 out of 17) exhibited coinciding peaks in both HR and SCL during amusement-driven scenes. Similar concurrent peaks in HR and SCL were also observed during scenes involving loud sounds, sudden reactions, and surprise events, indicating synchronized physiological responses to elevated emotional and sensory stimulation. About 18\% of the participants (3 out of 17) exhibited pronounced HR and SCL responses during narrative climactic moments. It was also observed from the SCL curves that nearly 47\% of the participants (8 out of 17) showed that SCL effectively captured emotional reactivity, with medium to high amplitude peaks corresponding to sudden reactions and humorous moments. In contrast, lower amplitude peaks were observed during empathy-driven scenes, reflecting comparatively lower but sustained emotional arousal. Furthermore, it was also observed that for 24\% of the participants (4 out of 17) there were little to no substantial change in overall SCL levels throughout the performance, indicating relatively stable physiological arousal. For some of these participants, one section of the performance appeared noticeably calmer than the other. Some of the participants also exhibited frequent fluctuations in heart rate (HR) during amusing yet auditory-stimulating scenes. These variations are likely associated with laughter-induced respiratory changes combined with dynamic emotional arousal, leading to continuous short-term modulation of HR during such segments. However, it was also found that certain participants demonstrated distinctive physiological responses in HR that differed from the overall such as:
\begin{enumerate}
    \item For some participants HR values dropped and lowered during sudden sounds and lights may be due to unexpected auditory stimuli.
    \item Some participants exhibited prolonged elevations in HR and SCL, likely due to heightened sensitivity to changes in lighting and sound or following a single intense scene that triggered sustained physiological activation.
\end{enumerate}

\subsection{Scene-wise physiological response}
This section presents a detailed analysis of physiological responses across the key scenes of the theatrical performance as mentioned in Table \ref{tab:keyscenes}. This helps to examine how HR and SCL vary in relation to specific narrative events, emotional contexts, and sensory stimuli within each scene. Table \ref{tab:variationkeyscenes} describes the scenes and corresponding variations in HR and SCL across all participants.
\begin{table}
\caption{Variation in HR and SCL during Key scenes}
\label{tab:variationkeyscenes}
\begin{tabular}{|p{1cm}|p{1.5cm}|p{2.5cm}|p{3.5cm}|p{3.5cm}|}
\hline
\textbf{Part} & \textbf{Scene} & \textbf{Expected emotion} &  \textbf{HR}  & \textbf{SCL}\\
\hline
Part 1 & Scene 1  & Amusement & Medium to high peaks & Low to high peaks \\
 & Scene 2  & Amusement & Low to high peaks & Low to high peak\\
 & Scene 3  & Empathy & Low to moderate peaks & Mostly stable \\
 & Scene 4  & Intense & Medium to high peaks & Low to medium peaks \\
 & Scene 5  & Amusement & Medium to high peaks & Low to high peaks \\
 & Scene 6  & Amusement & Medium range & Low to high peaks\\
 & Scene 7  & Sudden reaction/ Emotional arousal & Low to moderate peaks & High and medium peaks \\
\hline
Part 2 & Scene 1 & Amusement & High peaks & Low to medium peaks\\
 & Scene 2 & Amusement & Medium peaks & Low to high peaks\\
 & Scene 3 & Surprise/Intense reaction & Low to high peaks & High and medium peaks\\
 & Scene 4 & Amusement & Medium peaks & Low to medium peaks\\
 & Scene 5 & Amusement & Medium peaks & Low to medium peaks\\
 & Scene 6 & Sudden reaction & Stable & Medium to high peaks\\
 & Scene 7 & Surprise/ Sudden reaction & High peaks & Low to high peaks\\
 & Scene 8 & Empathy & Medium peaks & Low\\
\hline
\end{tabular}
\end{table}

From Table \ref{tab:variationkeyscenes}, it can be seen that for Part 1 of the performance amusement-based scenes consistently produced moderate to high HR peaks along with variable increases in SCL, indicating elevated audience engagement during humorous and loud moments. The empathy scene showed relatively low to moderate HR responses and stable SCL activity, suggesting lower autonomic arousal. The intense discussion scene produced moderate responses in both measures, while the sudden reaction scene elicited stronger SCL responses compared to HR, indicating higher sensitivity of SCL to abrupt stimuli. Similarly for Part 2 of the performance, physiological responses showed greater variation across scenes. Amusement scenes generally resulted in high HR peaks for some scenes, mostly those scenes also had loud sounds and reactions. However, for some amusement scenes there were moderate HR increases. Scenes involving surprise, intensity, and sudden reactions produced the highest physiological activation, with clear peaks in both HR and SCL. The sudden reaction scene showed stronger SCL activity compared to HR, indicating heightened autonomic response to abrupt events. The empathy scene resulted in moderate HR responses but low SCL activity, reflecting emotional engagement with low physiological arousal. 

\begin{figure}[htbp]
\centering
\begin{subfigure}[b]{0.8\textwidth}
    \centering
    \includegraphics[width=\textwidth]{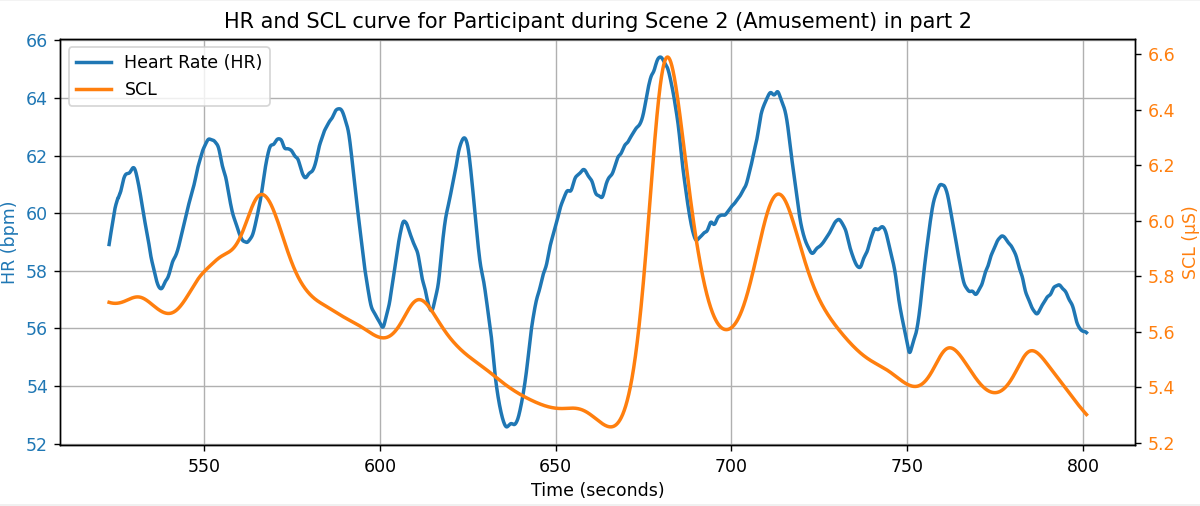}
    \caption{Amusement scene}
    \label{fig:HR_SCL_Amusement}
\end{subfigure}
\hfill
\begin{subfigure}[b]{0.8\textwidth}
    \centering
    \includegraphics[width=\textwidth]{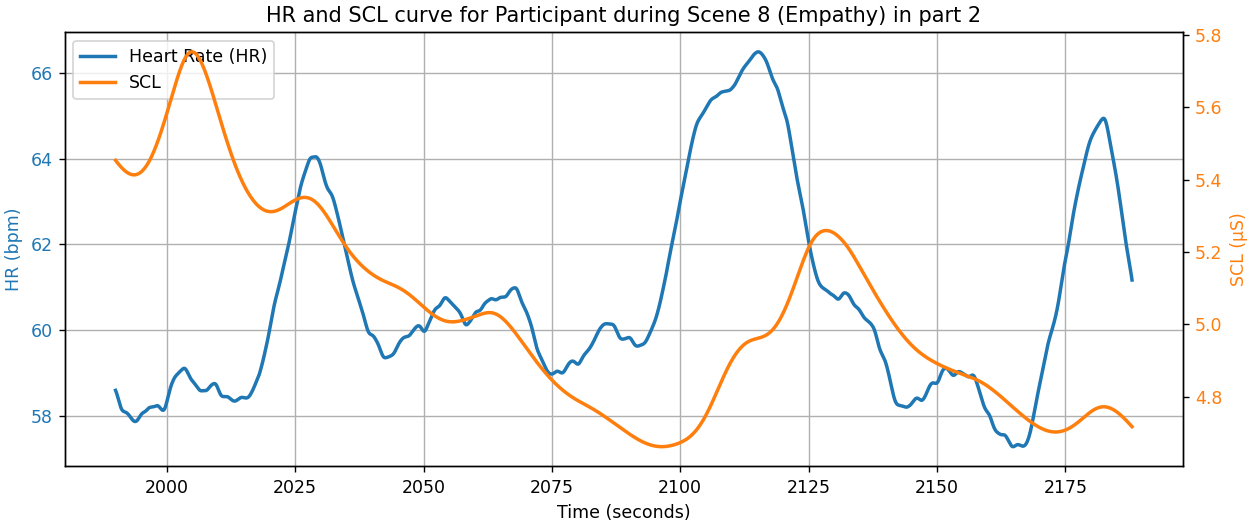}
    \caption{Empathy scene}
    \label{fig:HR_SCL_Empathy}
\end{subfigure}
\hfill
\begin{subfigure}[b]{0.8\textwidth}
    \centering
    \includegraphics[width=\textwidth]{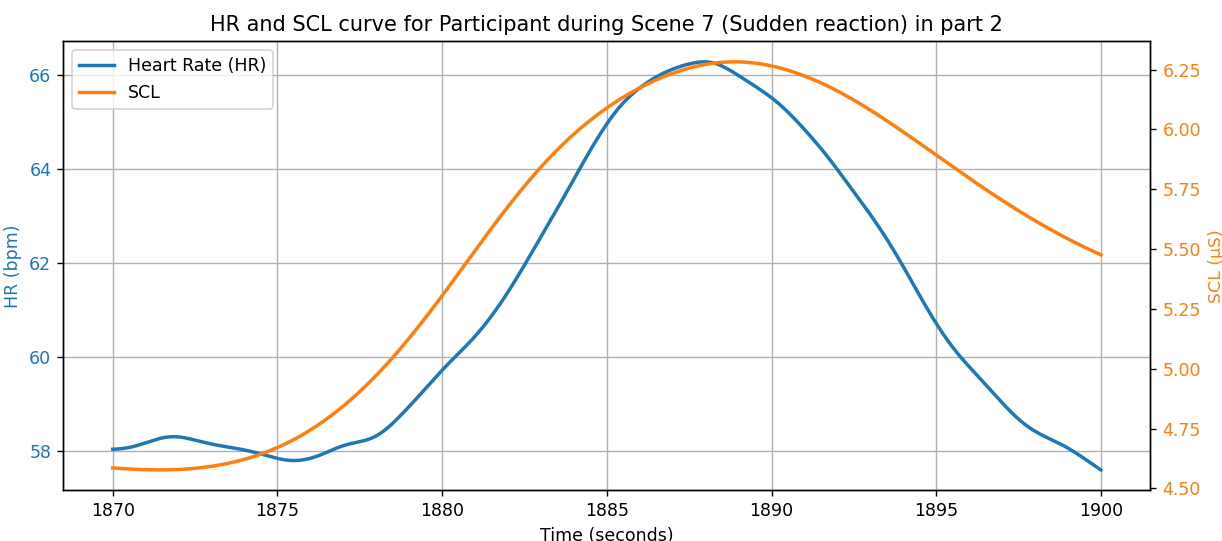}
    \caption{Sudden reaction scene}
    \label{fig:HR_SCL_SuddenReaction}
\end{subfigure}
\caption{Comparison of HR and SCL responses during amusement, empathy and sudden reaction scenes}
\label{fig:HR_SCL_Comparison}
\end{figure}

Figure \ref{fig:HR_SCL_Comparison} presents the HR and SCL responses of a particular participant during different scenes in Part 2 of the performance, including amusement, empathy, and sudden reaction scenes. As shown in Figures \ref{fig:HR_SCL_Amusement} and \ref{fig:HR_SCL_Empathy}, the amusement scene is characterized by greater variability in SCL, with values fluctuating from low to high levels as the scene progresses. In contrast, the empathy scene exhibits relatively low and stable SCL values with fewer peaks. HR responses remain within a moderate range in both scenes, showing noticeable fluctuations during the amusement scene and comparatively smaller variations during the empathy scene. However, during the sudden reaction scene (Figure \ref{fig:HR_SCL_SuddenReaction}), both HR and SCL exhibit a concurrent increase, resulting in a prominent physiological peak that indicates heightened arousal.

\subsection{Physiological inference and intensity analysis}
Following the participant-level and scene-wise analyses, a physiological inference and intensity analysis was conducted to examine the overall patterns and characteristics of physiological responses throughout the performance across participants. To facilitate interpretation, visualization techniques, including mean and standard deviation (SD) curves, z-score normalized heatmap, and intensity-based plots, were used to reveal common response patterns, peak, and variations in audience engagement across different segments of the performance. By relating these physiological responses to the time-aligned theatrical scenes, the analysis provides insights into the intensity and consistency of audience reactions during key emotional and dramatic moments.

\begin{enumerate}
    \item \textbf{Mean and standard deviation plots}
    Figure \ref{fig:HR_Mean_SD} shows mean HR and standard deviation values across participants for both parts of the theatre performance, indicating overall heart rate trend and variability in HR responses among participants. The HR values observed during Part 1 as shown in Figure \ref{fig:HR_Mean_SD_Part1} were generally higher than those recorded during Part 2 (Figure \ref{fig:HR_Mean_SD_Part2}). This suggests a greater level of physiological arousal during Part 1. Although both parts contained almost similar number of scenes with loud sounds and music and intense scenes, elevated HR in Part 1 may be attributed to increased emotional engagement. For example, participants may have experienced increased novelty, emotional engagement, or cognitive load during the initial phase of the performance. As the performance progressed, adaptation effects may have reduced the physiological response, resulting in comparatively lower HR values during parts despite similar auditory stimulation.
    
    Figure \ref{fig:SCL_Mean_SD} shows mean Skin Conductance Level (SCL) amplitude and standard deviation values across participants for both parts of the theatre performance, indicating overall SCL amplitude trend and variability in responses among participants. Subfigures \ref{fig:SCL_Mean_SD_Part1} and \ref{fig:SCL_Mean_SD_Part2} shows the mean and SD of SCL across participants for Part 1 and Part 2 respectively. In contrast to the HR responses, the SCL values were generally higher in Part 2 compared to Part 1. This suggests a stronger autonomic arousal during Part 2, particularly at the beginning of the segment. Concurrently, the HR values were also up during this period. The initial increase in SCL coincided with the onset of loud sounds and bright lighting, followed by an intense scene, which likely triggered a rapid sympathetic nervous system response. However, after this initial peak, SCL values gradually decreased and stabilized, becoming comparable to those observed in Part 1.
    \begin{figure}[htbp]
    \centering
    \begin{subfigure}[b]{0.9\textwidth}
        \centering
        \includegraphics[width=\textwidth]{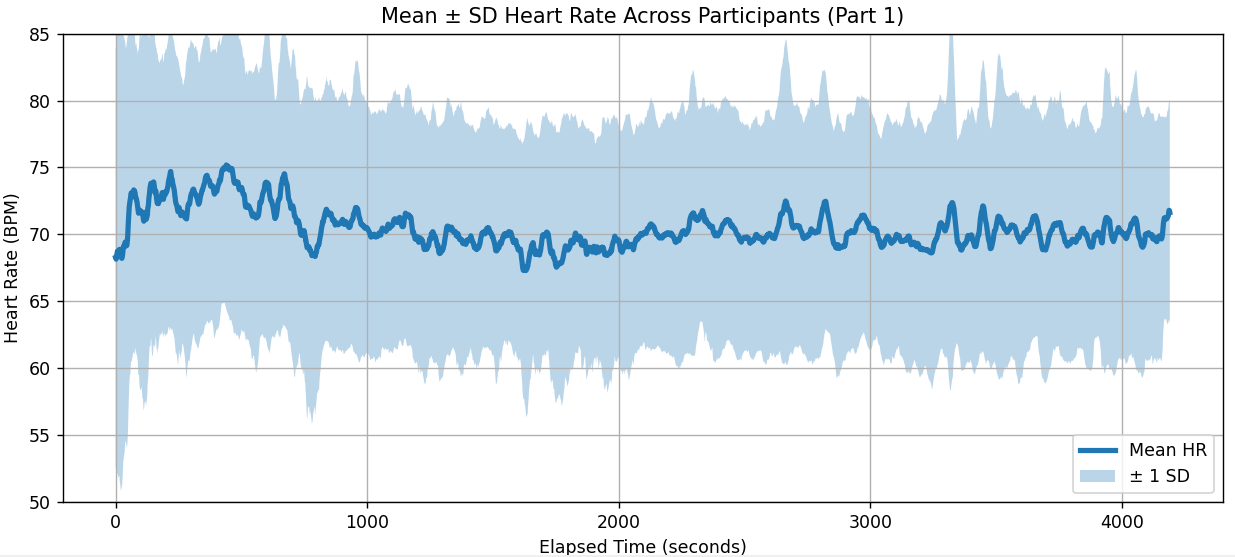}
        \caption{HR mean and SD curve for Part 1}
        \label{fig:HR_Mean_SD_Part1}
    \end{subfigure}
    \hfill
    \begin{subfigure}[b]{0.9\textwidth}
        \centering
        \includegraphics[width=\textwidth]{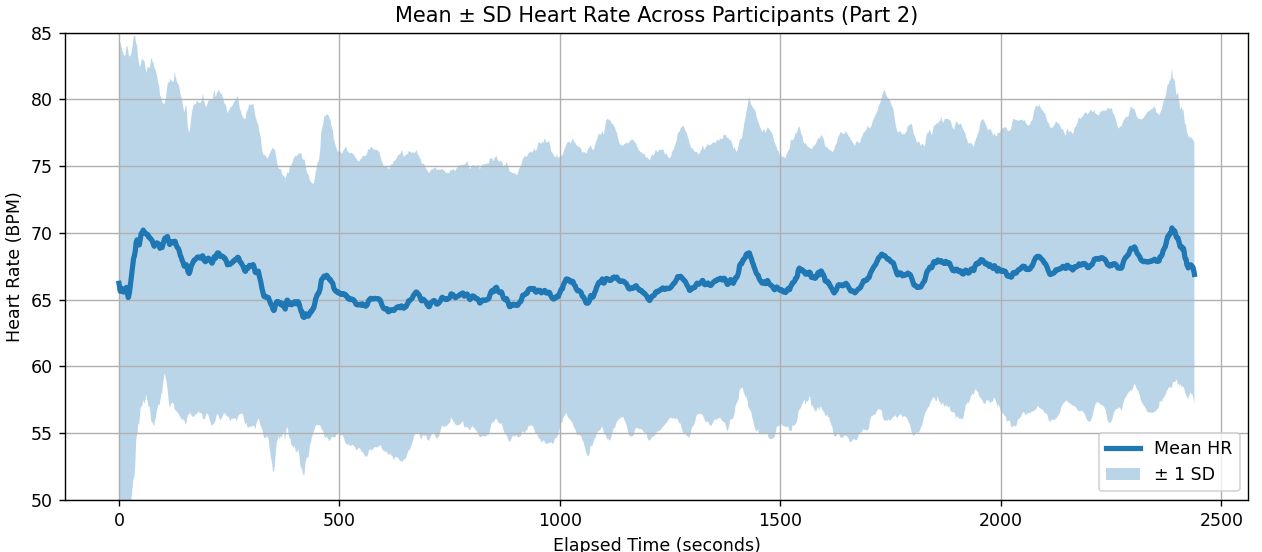}
        \caption{HR mean and SD curve for Part 2}
        \label{fig:HR_Mean_SD_Part2}
    \end{subfigure}
    \caption{Mean and SD curves for HR for Part 1 and Part 2 across participants}
    \label{fig:HR_Mean_SD}
    \end{figure}

    \begin{figure}[htbp]
    \centering
    \begin{subfigure}[b]{0.9\textwidth}
        \centering
        \includegraphics[width=\textwidth]{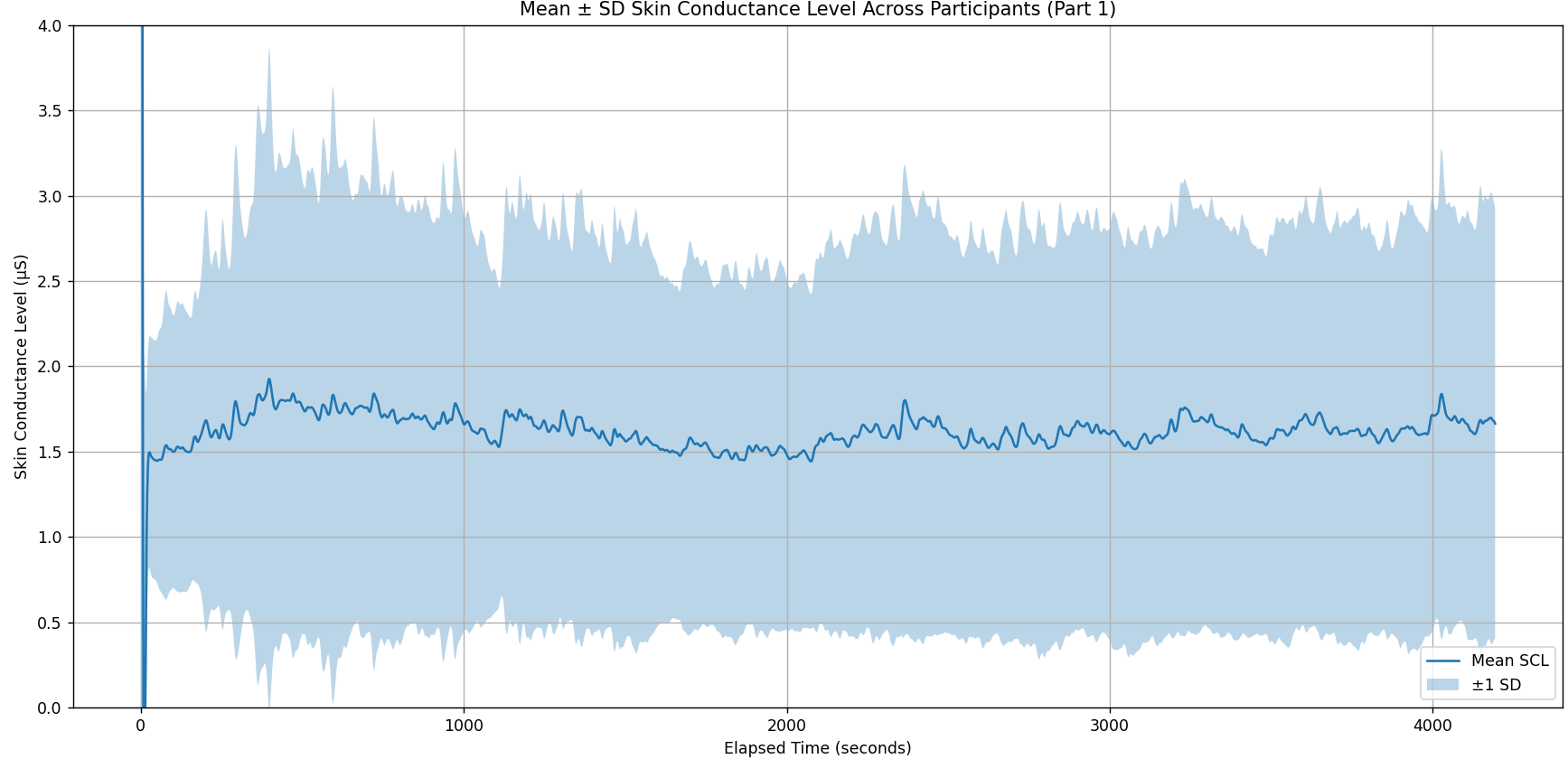}
        \caption{SCL mean and SD curve for Part 1}
        \label{fig:SCL_Mean_SD_Part1}
    \end{subfigure}
    \hfill
    \begin{subfigure}[b]{0.9\textwidth}
        \centering
        \includegraphics[width=\textwidth]{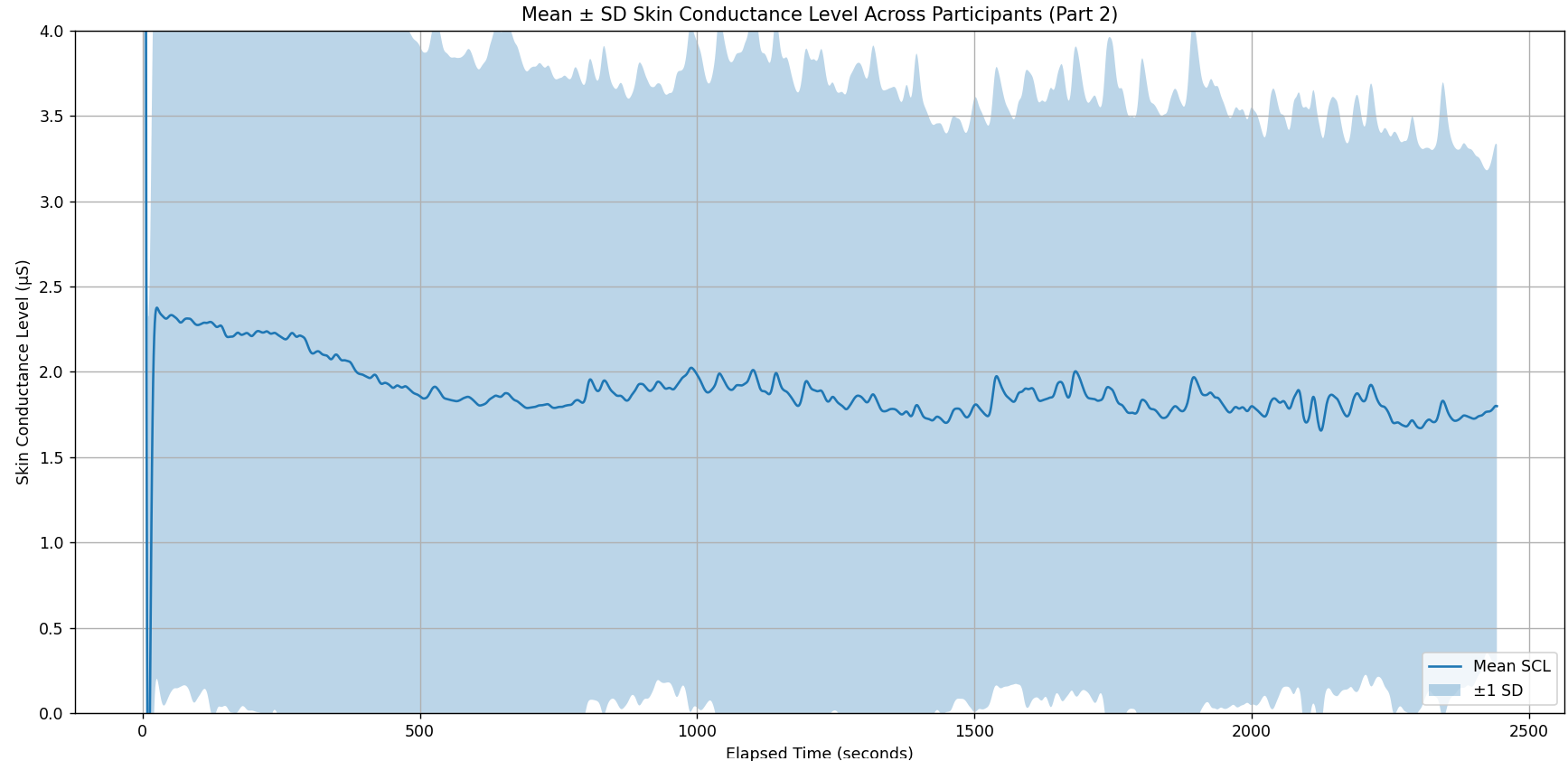}
        \caption{SCL mean and SD curve for Part 2}
        \label{fig:SCL_Mean_SD_Part2}
    \end{subfigure}
    \caption{Mean and SD curves for SCL for Part 1 and Part 2 across participants}
    \label{fig:SCL_Mean_SD}
    \end{figure}

    \item \textbf{Normalized z-score based heatmaps}
    \begin{figure}[htbp]
    \centering
    \begin{subfigure}[b]{\textwidth}
        \centering
        \includegraphics[width=\textwidth]{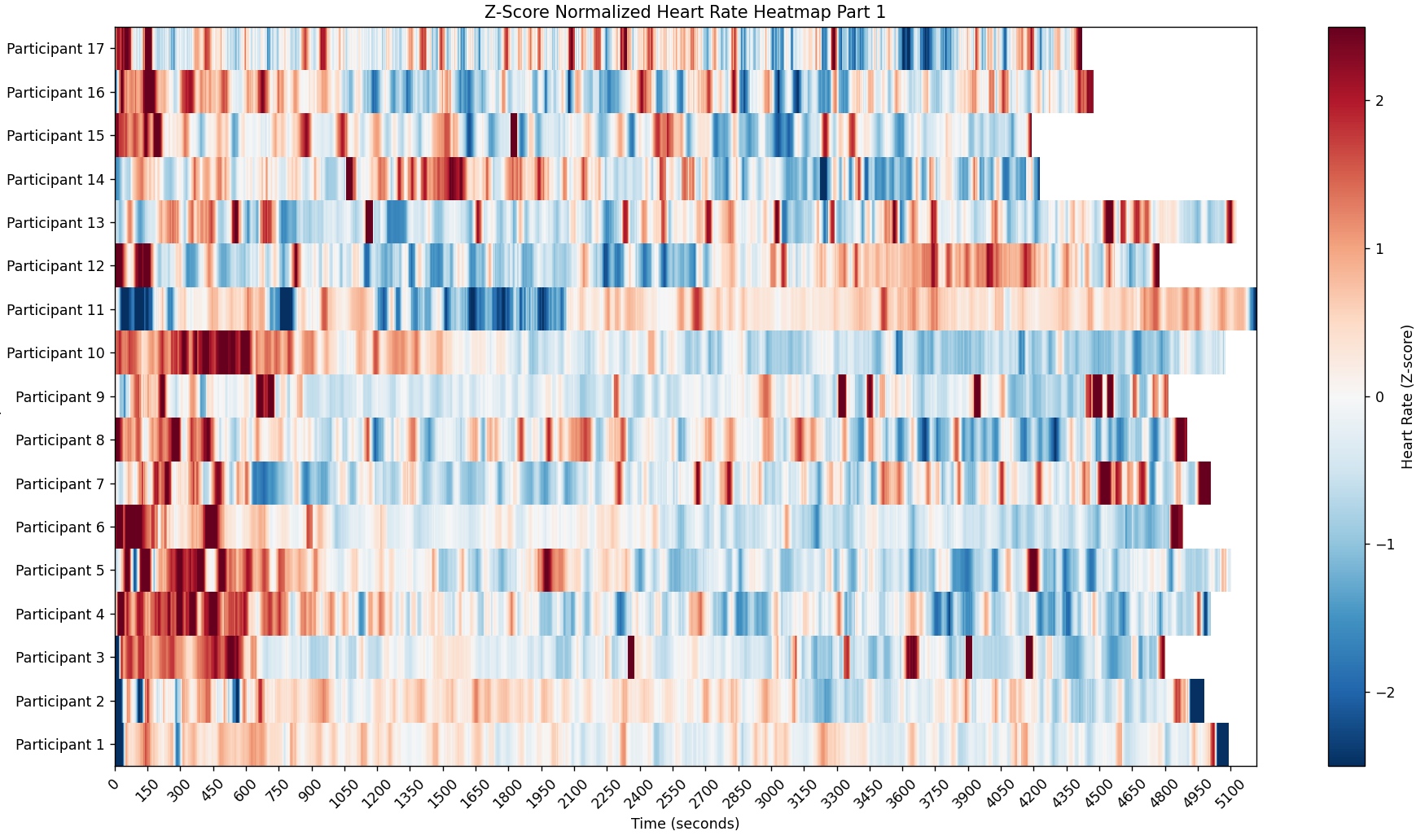}
        \caption{Heatmap for HR Part 1}
        \label{fig:HR_Heatmap_Part1}
    \end{subfigure}
    \hfill
    \begin{subfigure}[b]{\textwidth}
        \centering
        \includegraphics[width=\textwidth]{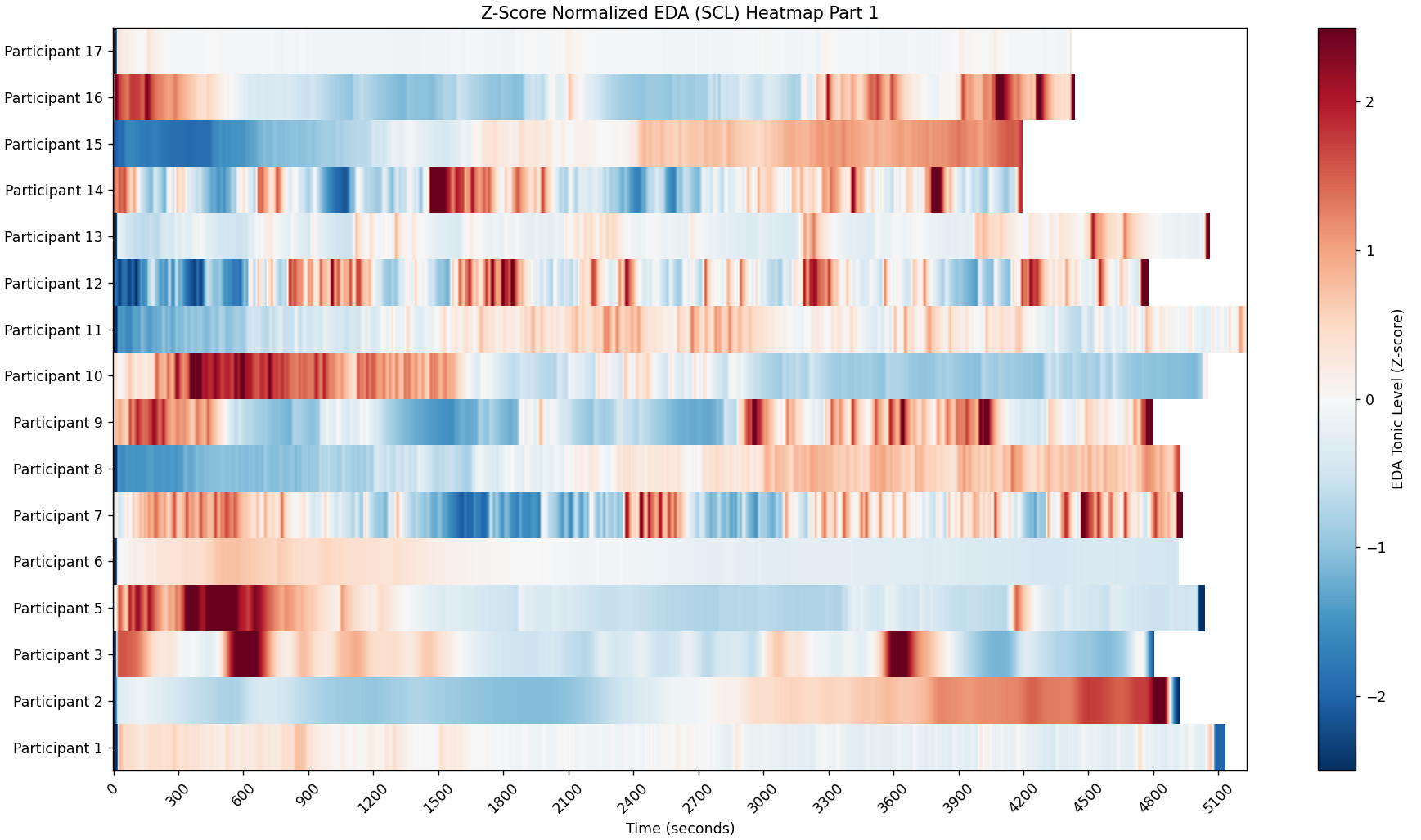}
        \caption{Heatmap for SCL Part 1}
        \label{fig:SCL_Heatmap_Part1}
    \end{subfigure}
    \caption{Heatmap for HR and SCL Part 1}
    \label{fig:Heatmap_HR_SCL_Part1}
    \end{figure}
    
    \begin{figure}[htbp]
    \centering
    \begin{subfigure}[b]{\textwidth}
        \centering
        \includegraphics[width=\textwidth]{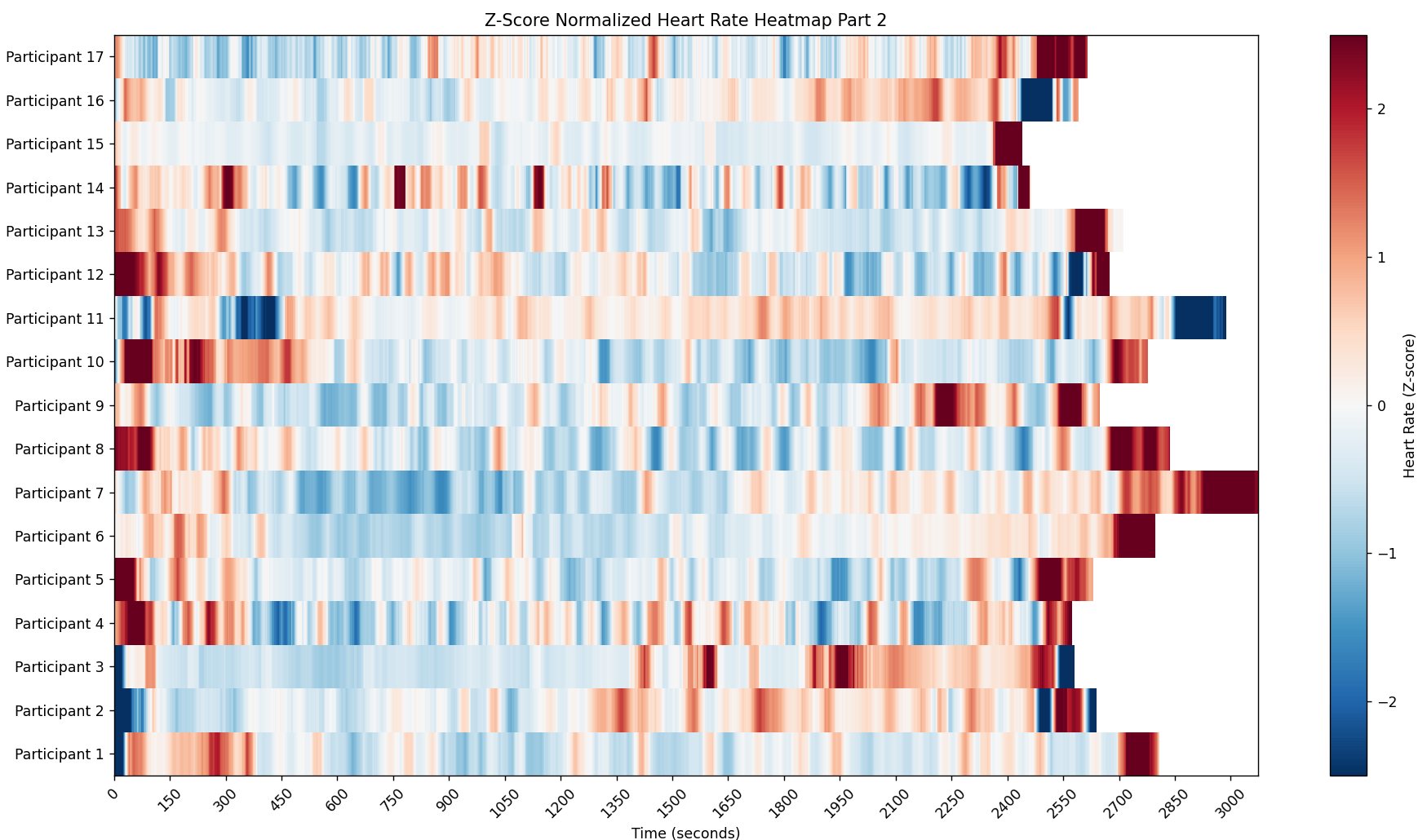}
        \caption{Heatmap for HR Part 2}
        \label{fig:HR_Heatmap_Part2}
    \end{subfigure}
    \hfill
    \begin{subfigure}[b]{\textwidth}
        \centering
        \includegraphics[width=\textwidth]{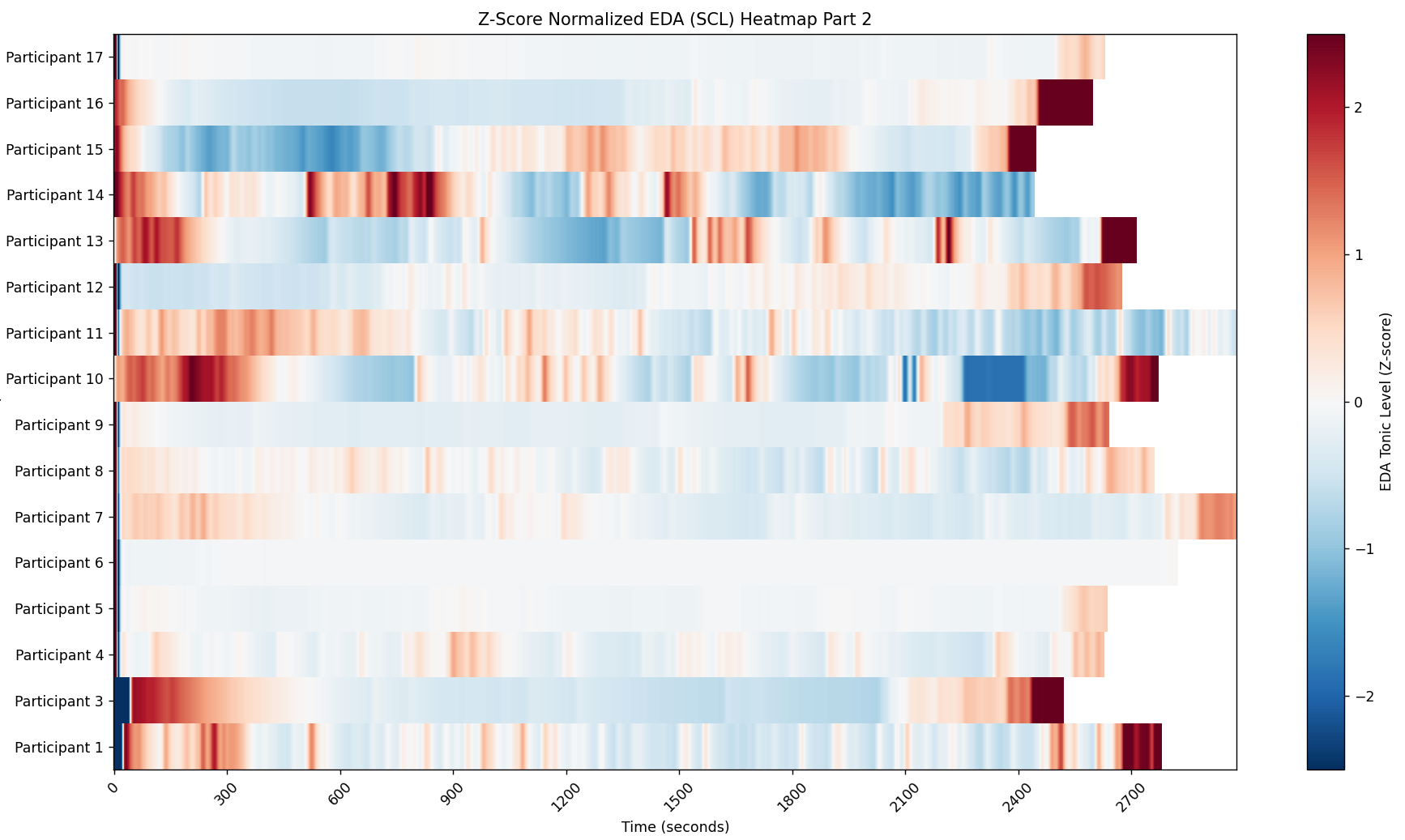}
        \caption{Heatmap for SCL Part 2}
        \label{fig:SCL_Heatmap_Part2}
    \end{subfigure}
    \caption{Heatmap for HR and SCL Part 2}
    \label{fig:Heatmap_HR_SCL_Part2}
    \end{figure} 
    Figures \ref{fig:Heatmap_HR_SCL_Part1} and \ref{fig:Heatmap_HR_SCL_Part2} shows the heatmap representation of z-scores for HR and SCL to provide a visual representation of physiological arousal patterns across participants during entire performance parts, Part 1 and Part 2 respectively. In the heatmaps, warmer colors represent positive deviations from baseline (higher-than-average arousal), while cooler colors indicate lower-than-baseline activity. These heatmaps are useful for identifying variable and concurrent patterns in HR and SCL along with participant-level variability in physiological responses.
    
    From Figure \ref{fig:Heatmap_HR_SCL_Part1}, it can be seen that HR and SCL heatmaps show a relatively consistent pattern of elevated arousal across a subset of participants, indicating a group-level response to the same segments. Several participants exhibit synchronised increases in both HR and SCL during the same time windows, as can be seen from the darker/warmer segments. In Part 2 as shown in Figure \ref{fig:Heatmap_HR_SCL_Part2}, the heatmaps show clearly synchronised instants of arousal between HR and SCL prominantly at the beginning, during intense scene and around narrative climax. Some participants exhibit elevated SCL with relatively stable HR, while others show the opposite trend. This suggests more individualized responses during Part 2, possibly due to differences in emotional interpretation. Overall, the comparison between HR and SCL across both parts indicates that Part 1 is characterised by more consistent and shared arousal responses across participants, whereas Part 2 shows greater inter-individual variability. 

    \item \textbf{Intensity analysis}
    To examine the overall physiological arousal throughout the theatre performance, an intensity index was calculated by averaging the z-scored HR and SCL. The resulting intensity measure provides a unified representation of autonomic arousal. Intensity was analysed using line graphs to show the variation in intensity across different segments within Part 1 and Part 2. Figure \ref{fig:Intensity_plot} shows the intensity vs time plot for Part 1 and Part 2 of the performance for key scenes (as mentioned in Table \ref{tab:keyscenes}). In these intensity curves, emotional labels are used in the legend instead of segmented key scenes. This was done because the performance structure does not always align perfectly with clearly separated scene boundaries between all the participants, leading to minor temporal deviations in the annotated timestamps. Therefore, representing the data in terms of emotional states provides better interpretation of physiological intensity patterns across the performance.
    
    As shown in Figure \ref{fig:Intensity_plot_part1}, for Part 1, the highest intensity values were observed during the empathy scene, indicating strong audience engagement and physiological arousal. Increased intensity was also observed during several amusement scenes. However, among amusement scenes, initial amusement scene which involved loud noise and scream caused more arousal. Overall, both empathy and amusement contributed to audience arousal, with the empathy scene producing the most significant response. The intensity values in Part 2, as can be seen from Figure \ref{fig:Intensity_plot_part2}, were lower than those observed in Part 1. However, in Part 2 several peaks were identified during amusement scenes and during scenes that combined entertaining elements with loud reactions. Elevated intensity was also observed during the empathy scene associated with the narrative climax, suggesting strong emotional engagement with the storyline. Although the overall intensity was reduced compared to Part 1, these findings indicate that both emotionally significant and highly engaging scenes continued to evoke audience responses throughout Part 2 as well.

    \begin{figure}[htbp]
    \centering
    \begin{subfigure}[b]{\textwidth}
        \centering
        \includegraphics[width=1\textwidth]{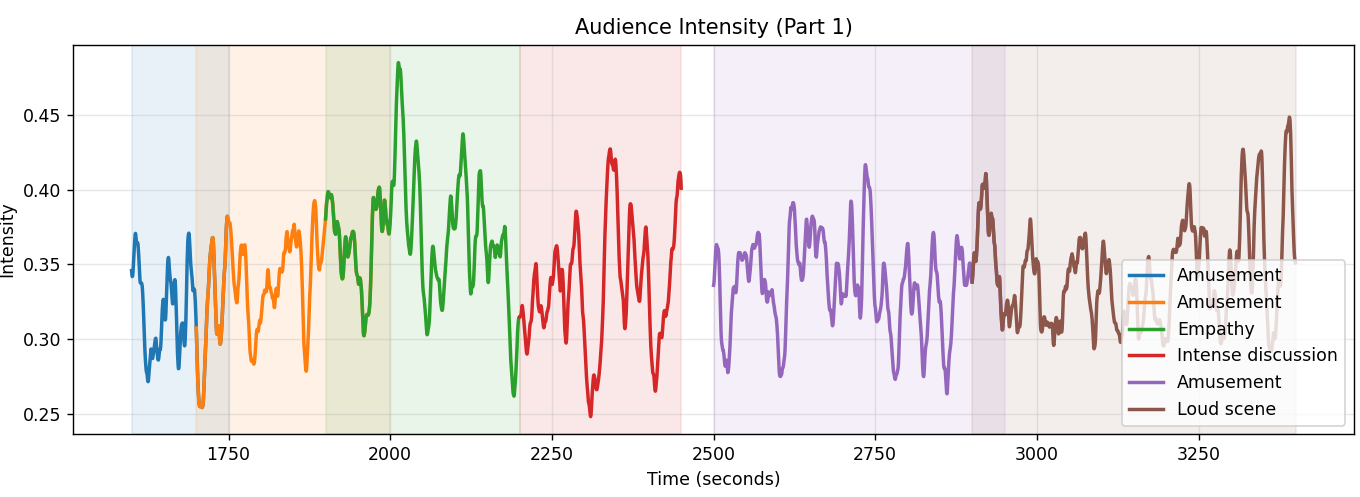}
        \caption{Intensity vs time plot-Part 1}
        \label{fig:Intensity_plot_part1}
    \end{subfigure}
    \hfill
    \begin{subfigure}[b]{\textwidth}
        \centering
        \includegraphics[width=1\textwidth]{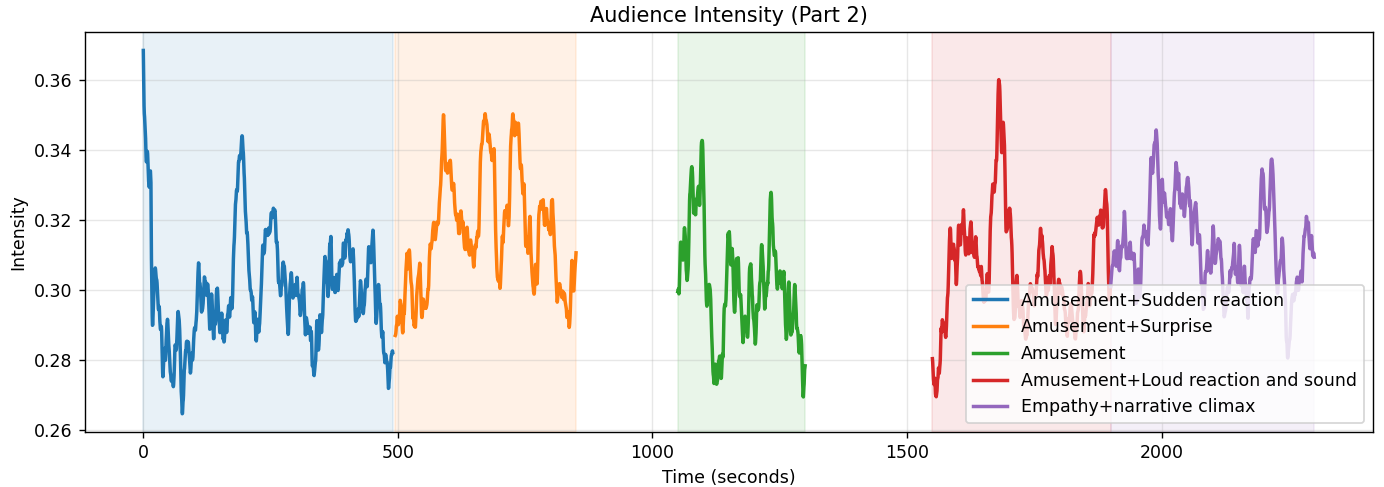}
        \caption{Intensity vs time plot-Part 2}
        \label{fig:Intensity_plot_part2}
    \end{subfigure}
    \caption{Intensity analysis for Part 1 and Part 2 across different scenes}
    \label{fig:Intensity_plot}
    \end{figure}
    
\end{enumerate}

\section{Discussion}
This study explored how physiological responses change during real-life recreational event, in this case live theatrical performance. To identify the patterns or changes in physiological signals, we recorded data from 17 participants. The signals include ECG, and GSR signals recorded from Shimmer sensing device. We conducted a demographic analysis to identify age-wise and gender-wise distribution of participants. We conducted a comprehensive visual analysis by extracting features such as HR and SCL from ECG and GSR signals respectively. In addition to the visual analysis, we also conducted participant-level observations and scene-wise physiological signal analyses. Participant-level observations were used to identify individual response patterns and common trends across the performance, while the scene-wise analysis examined physiological responses associated with specific theatrical moments. 

The results of these analysis indicate that different theatrical moments evoked distinct patterns of physiological activation, demonstrating the potential of physiological measures for understanding audience engagement. Participant-level observations revealed that there were coinciding peaks for HR and SCL in various scenes involving amusement and sudden reaction for most of the participants. Furthermore, the scene-wise physiological responses demonstrated differences among humorous, empathy-driven, intense, and sudden reaction scenes, specifically for SCL. It also revealed that amusement scenes generally elicited moderate to high physiological responses, particularly when accompanied by loud sounds, or sudden reactions. Empathy-related scenes were characterized by comparatively lower and more stable responses, with fewer pronounced peaks. In contrast, scenes involving surprise, intense reactions, or dramatic events produced more physiological activation, often reflected by concurrent increases in both HR and SCL. Further analysis using mean and SD curves over the entire performance revealed differences in physiological responses between Part 1 and Part 2. In particular, the analysis showed which part of performance exhibited higher average HR and SCL values, providing insights into the relative intensity of audience engagement throughout the performance. In addition to this, we also explored the normalized z-scores and plotted the heatmaps to see the variable and concurrent segments among participants throughout the performance part. The intensity analysis conducted using z-scores, further demonstrated that physiological activation was concentrated around specific emotional and dramatic events. 

Together, these analyses provided a comprehensive understanding of participant engagement and physiological activation throughout the performance. The combined analysis of HR and SCL offers valuable insights and highlights the potential of physiological sensing as a tool for evaluating responses in live performance settings. Furthermore, this approach has practical applicability in domains where understanding real-time emotional engagement is essential such as theatre design, audience experience research, affective state analysis, interactive performance evaluation and development of recommendation system for interventions \cite{RecommendationSystem,RecommendationSystem1,AffectiveStateAnalysisDiscussion}. 

While the present study provides preliminary evidence of variations in physiological responses across different theatrical contexts, more comprehensive quantitative analysis is required to validate and further characterize these patterns. Preliminary statistical analyses have been conducted; however, these require further investigation before robust conclusions can be drawn. Future work will therefore focus on more rigorous statistical evaluation and exploratory clustering of physiological and applying machine learning based analysis to identify and predict meaningful physiological patterns. This will enable automated detection of audience arousal states and deeper interpretation of engagement so that this could facilitate real-time or automated detection of engagement patterns and support adaptive interventions.

\bibliographystyle{splncs04}
\bibliography{Bibliography}
\end{document}